# Global Warming and Caspian Sea Level Fluctuations

Reza Ardakanian[1], Seyed Hamed Alemohammad[2]

[1]*Assisstant Professor, Department of Civil Engineering, Sharif University of Technology, P.O. Box: 11155-9313, Tehran, Iran, Email: ardakanian@civil.sharif.edu*

[2] *Graduate Student, Department of Civil Engineering, Sharif University of Technology, P.O. Box: 11155-9313, Tehran, Iran, Email: alemohammad@civil.sharif.edu*

## Abstract

Coastal regions have a high social, economical and environmental importance. Due to this importance the sea level fluctuations can have many bad consequences. In this research the correlation between the increasing trend of temperature in coastal stations due to Global Warming and the Caspian Sea level has been established. The Caspian Sea level data has been received from the Jason-1 satellite. It was resulted that the monthly correlation between the temperature and sea level is high and also positive and almost the same for all the stations. But the yearly correlation was negative. It means that the sea level has decreased by the increase in temperature.

## 1- Introduction

After the Industrial Revolution the emission of green house gases to the atmosphere has increased by a significant rate. According to the statistics, the amount of Carbon Dioxide, as one of the most important green house gases, has increased from 280 ppm in 1750 to 379 ppm in 2005 **[1]**. Of course this increase has had a bigger rate in the latest decades. Due to this increase the global temperature has been risen which results in changing the stationary trend of the climate variables of the earth. This phenomena has been entitled *"Climate Change"*. In recent years many researches has been done to determine the effects of Climate Change on water cycle and it's components like precipitation, snow and etc. One of the most important effects of climate change that has been discovered is the changes that result in sea level.

The IPCC[1] 4th Assessment Report states that "Present-day sea level change is of considerable interest because of its potential impact on human populations living in coastal regions and on islands". The current rate of sea level rise has been reported to be 1-2.5 mm/yr in the 20th century **[2]**. Modern satellite measurements reveal that since 1993, sea-level has been rising at an average rate of about 3 mm/yr, substantially faster than the average for the 20th century of about 1.7 mm/yr, estimated from coastal sea-level measurements.

The main contributions to the 20th and 21st century sea-level change are:

- Thermal expansion of the oceans (water expands as it warms),
- The addition of mass to the oceans from the melting of glaciers and ice caps in regions like Himalayas, Alaska, Patagonia, etc.,
- The exchange of mass with the ice sheets of Antarctica and Greenland,
- The exchange of mass with terrestrial water storages (groundwater, aquifers, dams, lakes)

Caspian Sea is the biggest lake in the world. It is located between the Europe and South West of Asia. Iran, Azerbaijan, Kazakhstan, Russia and Turkmenistan are the neighbors of the sea. As Caspian Sea has many sources of Oil and Gas it has a very economical and environmental importance. Due to the importance of the sea its fluctuations will be considerable.

M.R. Meshkani and A. Meshkani (1999) modeled the Caspian Sea Level Fluctuations using a stochastic modeling. In their research a dynamic correlation has been established between the precipitation and temperature in Bandar-e-Anzali station and the Sea Level. But no consideration has been made to the trends of he temperature due to global warming. **[3]**

Naidenov (2002) has used a nonlinear model to predict the variations of the Caspian Sea level change. The seawater budget has been described by a system of nonlinear stochastic differential equations **[4]**. Loehle (2004) has analyzed two 300-year temperature series to detect the change in temperature. Seven models have been fitted to the series which hadn't contained the 20th century data. The projections of the six of the models have shown a warming trend over the 20th century similar in timing and magnitude to the Northern Hemisphere instrumental series **[5]**.

Elguindi and Giorgi (2007) has simulated the Caspian Sea Level Fluctuations using the results of the outputs of the Regional Climate Model. They have considered the A2 scenario as the situation in 2071 – 2100 and the sea level has been predicted using the Regional Climate Model resulted form the Global Climate Model. They

---

[1] *International Panel of Climate Change (IPCC)*

have concluded that the sea level will decrease much more than that of the 20$^{th}$ Century. **[6]**

In this research we are going to search on a correlation between the sea level data and the temperature data at coastal gauges. The necessary sea level data has been accessed through the Jason-1sattelite. The temperature data at 7 coastal stations has been received from the National Meteorological Organization of Iran. As the satellite data are available from September 1992 and the temperature data are available up to 2005, the period of September 1992 to December 2005 has been selected for the analysis period.

In the following a short description of the Jason-1 satellite and also the working procedure in addition to the conclusion will be presented.

**2- Data Description**

Use of satellites to measure the sea levels has more and more increased in recent decades. The U.S. Department of Agriculture's Foreign Agricultural Service[2], in co-operation with the National Aeronautics and Space Administration[3], and the University of Maryland, are routinely monitoring lake and reservoir height variations for approximately 100 lakes located around the world. This project is unique, being the first of its kind to utilize near-real time radar altimeter data over inland water bodies in an operational manner. Surface elevation products are produced via a semi-automated process. The project utilizes near-real time radar altimeter data from the Poseidon-2 instrument on-board the Jason-1 satellite which was launched in December, 2001. In addition, historical archive data is used from the TOPEX/POSEIDON mission (1992-2002).

The data available in this project are relative lake height variations computed from TOPEX/POSEIDON (T/P) and Jason-1 altimetry with respect to a 10 year mean level derived from T/P altimeter observations. In this research the Caspian Sea level data has been used from this project to decrease the possible errors in the tide gauge measurements. Figure 1 shows the monthly Caspian Sea level fluctuations from September 1992 to December 2005 driven from 10-day Jason-1 data.

In the period of 1992-2005 there were 8 months that have two series of data: one from the TOPEX/POSEIDON and the other form the Jason-1. Data used for these months, which contain January 2002 to August 2002, are the averages from both satellites.

---

[2] *USDA-FAS*
[3] *NASA*

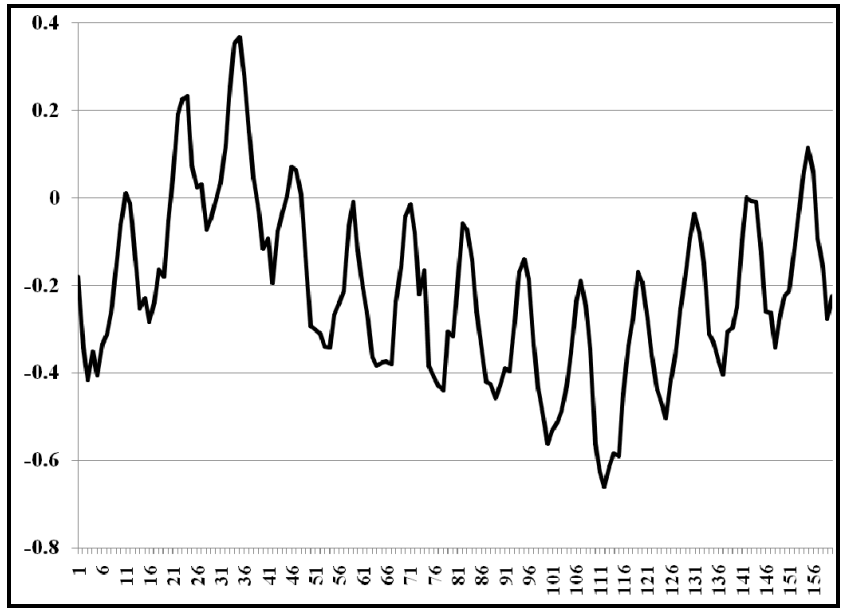

Fig 1 – Monthly Caspian Sea level fluctuations with respect to the 10 year mean

The temperature data has been used in seven stations namely: Astara, Bandar-e-Anzali, Rasht, Ramsar, Noushahr, Babolsar and Gorgan. The data has been derived form the yearly reports of the National Meteorological Organization of Iran. As an example the yearly series of temperature at Astara station from 1986 to 2005 has been shown in figure 2.

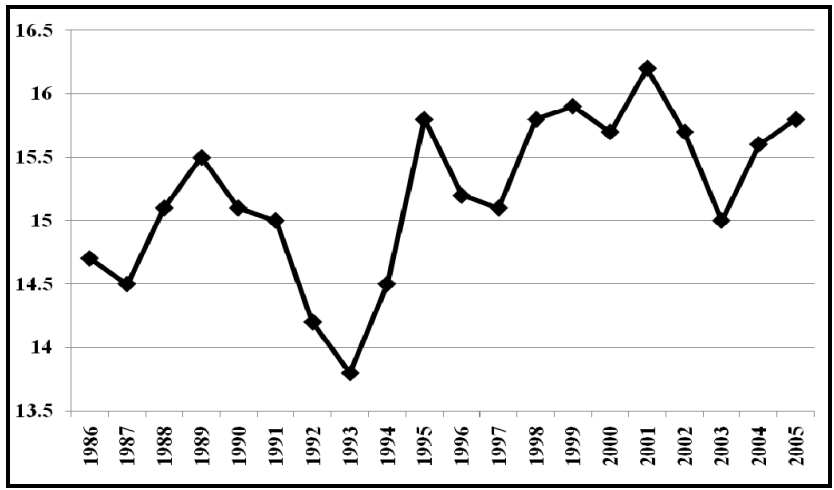

Fig 2 – Yealry temperature at Astara station

### 3- Data Analysis

Two main steps are presented here: First, detecting the trend in temperature time series of the stations, second, detecting the correlation between the temperature and sea level.

#### 3-1- Trend Detection

To detect the trend in a time-series we will use two methods:

1. Cumulative Deviation Test
2. Regression Analysis

Each of these methods will be described in the following:

**1- Cumulative Deviation Test**

This test is defined on the basis of the cumulative deviations from the mean, as it follows:

$$S_k = \sum_{i=1}^{k}(Y_i - \bar{Y}), \quad k = 1,...,n \tag{1}$$

in which:

$Y_i$: are the time-series data

$\bar{Y}$: is the mean of $Y_i$

$n$: is the number of data

then the $S^*_k$ will be defined such as:

$$S^*_k = S_k / D_Y \tag{2}$$

$$D_Y^2 = \sum_{i=1}^{n}(Y_i - \bar{Y})^2 / n \tag{3}$$

According to the value of $S^*_k$ the Q parameter which is the sensitivity to the deviation from the mean is defined as:

$$Q = \max_{0 \leq k \leq n} |S^*_k| \tag{4}$$

For 95% confidence interval the Q parameter should be less than 1.27 so that we can say there is no trend.

2- **Regression Analysis**

In this test a simple regression equation will be calculated for the series in the form:

$$Y = a + bX \tag{5}$$

In which:

Y is the dependent variable (ex. sea level) and X is the independent variable (ex. temperature)

a and b are the coefficients of the regression which will be calculated using the least square method.

To test the significance of the b (slope of the trend) the T parameter should be calculated as below:

$$T = \frac{b}{\sqrt{MSE/S_{xx}}} \tag{6}$$

$$S_{xx} = \sum_{i=1}^{n}(X_i - \bar{X})^2 \tag{7}$$

in which

MSE is the Mean Square of the Errors.

The absolute value of the calculated value of the T should be compared to $t_{\alpha/2, n-2}$, if $|T| < t_{\alpha/2, n-2}$ then the slope is significant.

3-2- **Correlation Detection**

The second analysis has been made to detect the correlation between the temperature and sea level. In this regard the correlation coefficient has been calculated using (8).

$$r_{zy} = \frac{n\sum_{i=1}^{n} z_i y_i - \sum_{i=1}^{n} z_i \sum_{i=1}^{n} y_i}{\sqrt{n\sum_{i=1}^{n} z_i^2 - \left(\sum_{i=1}^{n} z_i\right)^2} \sqrt{n\sum_{i=1}^{n} y_i^2 - \left(\sum_{i=1}^{n} y_i\right)^2}} \tag{8}$$

In which z and y are two time series (ex. temperature and sea level).

The correlation has been calculated for monthly, seasonal and yearly data. It means three different correlation coefficients has been calculated for each station, one for the monthly temperature data and monthly sea level, the other for the seasonal temperature data and seasonal sea level and the last one for the yearly temperature data and yearly sea level.

For the seasonal data the normal seasons of the year (Spring, Summer, Autumn and Winter) has been considered.

**4- Results**

The results are divided to two parts, first the results of the trend tests and second the results of the correlation coefficient.

Table 1 shows the results of the trend tests. In the Regression Analysis the confidence interval to clculate $t_{\alpha/2,n-2}$ has been chosen to be 95%. Also, Figure 3 shows three sample stations data with the trend line plotted on the monthly temperature data.

Table 1- The results of the trend tests for the temperatuare

| Station | Cumulative Deviation Test | | Regression Analysis | | | |
|---|---|---|---|---|---|---|
| | Q | Trend | b | T | t | Trend |
| Astara | 11.7848 | Yes | 0.061 | 0.334108 | 2.10092 | Yes |
| Babolsar | 22.9002 | Yes | 0.026 | 0.247812 | 2.009 | Yes |
| Banda-e-Anzali | 14.3718 | Yes | 0.0009 | 0.006114 | 2.009 | Yes |
| Gorgan | 8.3122 | Yes | 0.001 | 0.007142 | 2.009 | Yes |
| Noushahr | 16.1008 | Yes | 0.031 | 0.189233 | 2.052 | Yes |
| Ramsar | 18.7054 | Yes | 0.013 | 0.096041 | 2.009 | Yes |
| Rasht | 14.2665 | Yes | 0.027 | 0.210979 | 2.009 | Yes |

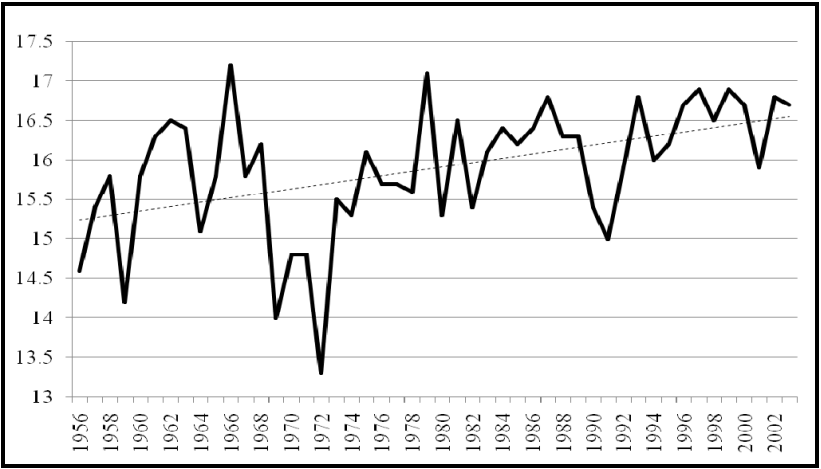
*Rasht Station*

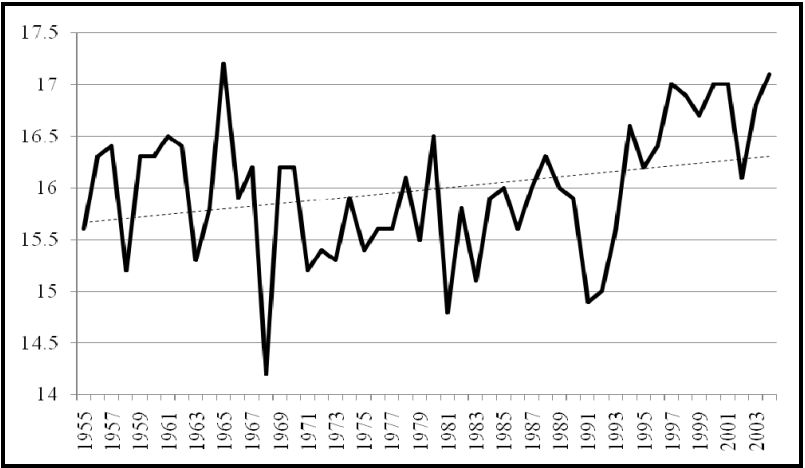
*Ramsar Station*

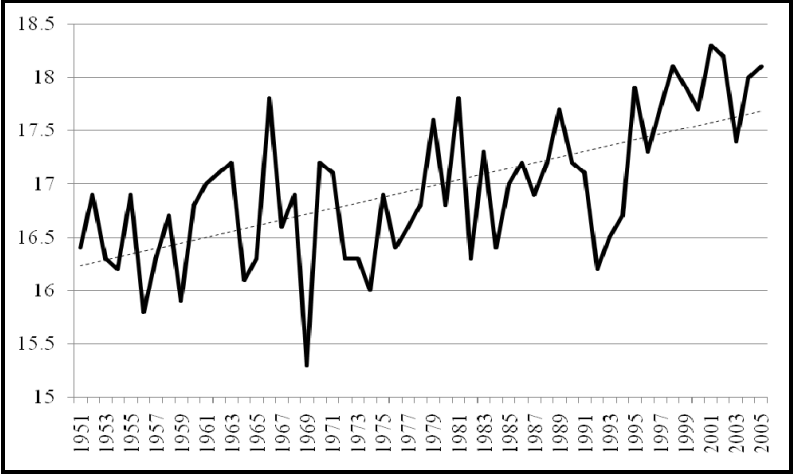
*Babolsar Station*

Fig 3 – Yearly time series of temperatuare with their trend line

The second part of the results contain the correlation coefficients between the sea level elevation and the temperatuare at different stations. As It was mentioned before the correlations coefficient has been caclculated for monthly, seasonal and yearly series for each station. Table 2 shows the results.

Table 2 – The correlation coefficients between the tempreatuare and sea level

| Station | Monthly Correlation Coefficients | Seasonal Correlation Coefficients | Yearly Correlation Coefficients |
|---|---|---|---|
| Astara | 0.5653966 | 0.5414004 | -0.2876049 |
| Banda-e-Anzali | 0.5539511 | 0.5285167 | -0.3135678 |
| Rasht | 0.5666350 | 0.3880966 | -0.1840101 |
| Ramsar | 0.5465243 | 0.3355391 | -0.3404678 |
| Noushahr | 0.5577893 | 0.3554810 | -0.3491657 |
| Babolsar | 0.5602099 | 0.5389376 | -0.3793325 |
| Gorgan | 0.5609775 | 0.37606941 | -0.4816182 |

5- **Conclusion**

As it was shown in the last section all the stations have trend in their yearly temperature series, according to the results of the both tests. This is an indication of the Climate Change and shows the need for more assessment and research.

The correlation coefficient for the monthly series in all station was positive, high and almost the same amount. But, although the seasonal correlation was positive, it wasn't the same for all stations. The important thing is that the yearly correlation coefficients were all negative. This means that although the yearly trend in the temperature was positive (increasing) the yearly trend in sea level is negative (decreasing). Finally we can say that an increase in the yearly temperature results in decrease of the sea level. Of course the decreasing trend can be detected through the regression analysis in the sea level data.

The Caspian Sea Level Fluctuations result from many parameters that temperature is one of them and a study of the correlation between the sea level and temperature alone can't determine the trend in sea level. It's worth mentioning that the monthly correlation coefficients had a high positive value. It's suggested to consider other parameters such as precipitation, air pressure, etc for modeling the trend in sea level data.

## 6- References

[1] IPCC (2007). "Climate Change 2007: The Physical Science Basis. Contribution of Working Group I to the Fourth Assessment Report of the Intergovernmental Panel on Climate Change" [Solomon, S., D. Qin, M. Manning, Z. Chen, M. Marquis, K.B. Averyt, M. Tignor and H.L. Miller (eds.)]. Cambridge University Press, Cambridge, United Kingdom and New York, NY, USA, 996 pp.

[2] Gornitz, V. (2000). "Impoundment, groundwater mining, and other hydrologic transformations: Impacts on global sea level rise." *Sea Level Rise: History and Consequences* (B.C. Douglas, M.S. Kearney, and S.P. Leatherman, Eds.), p. 97-119.

[3] Meshkani, M.R. and Meshkani, A., (1997) "Stochastic modeling of the Caspian sea level fluctuations", Theoretical and Applied Climatology, 58, pp. 189-195.

[4] Naidenov, V.I. and Shveikina, V.I. (2002). "A Nonlinear Model of Level Variations in the Caspian Sea." *J. of Water Resources*, 29 (2), p. 160-167.

[5] Loehle, C. (2004). "Climate change: detection and attribution of trends from long-term geologic data." J. of Ecological Modelling, 171, p. 433-450.

[6] Elguindi, N. and Giorgi, F., (2007) "Simulating future Caspian sea level changes using regional climate model outputs", Clim. Dyn., 28, pp. 365-379.

[7] Radar altimeter data from the NASA/CNES Topex/Poseidon and Jason-1 satellite missions. Time series of altimetric lake level variations from the USDA Reservoir Database at:
http://www.pecad.fas.usda.gov/cropexplorer/global_reservoir

[8] Marc F.P. Bierkens, *Stochastic Hydrology*, Utrecht University, Utrecht, The Netherlands, 2007.